\documentclass[nofootinbib,floats,floatfix,eqsecnum,prd,aps]{revtex4}
\usepackage[latin1]{inputenc}
\usepackage[T1]{fontenc}
\usepackage[dvips]{graphicx}
\usepackage{hyperref}
\usepackage{dcolumn,epsfig,minitoc}
\usepackage{amssymb,amsmath}
\usepackage[usenames]{color}
\def\a{\alpha}
\def\b{\beta}

\def\e{\epsilon}
\def\f{\phi}
\def\g{\gamma}

\def\k{\kappa}
\def\l{\lambda}
\def\m{\mu}

\def\t{\tau}

\def\G{\Gamma}

\def\G{\Gamma}

\def\l{\lambda}

\def\exp#1{{\rm e}^{#1}}

\def\o1{$O_{1}$}
\def\o2{$O_{2}$}

\def\lb{\label}
\newcommand{\beq}{\begin{equation}}
\newcommand{\eeq}{\end{equation}}
\newcommand{\bea}{\begin{eqnarray}}
\newcommand{\eea}{\end{eqnarray}}

\begin{document}

\title{Dark Energy from holographic theories  with 
hyperscaling violation }

\author{Mariano Cadoni$^{1,2}$ and Matteo Ciulu$^{1}$}
\affiliation{$^{1}$ Dipartimento di Fisica, Universit\`a di 
Cagliari.\\ $^{2}$ INFN, Sezione di
Cagliari.\\ Cittadella Universitaria, 09042 Monserrato, Italy.}

\date{\today }

\begin{abstract}
We show that  analytical continuation  maps   scalar solitonic 
solutions of Einstein-scalar 
gravity, interpolating 
between an  hyperscaling violating and an Anti de Sitter (AdS) region, in  flat FLRW 
cosmological  solutions 
sourced by a scalar field.  We generate in this way exact 
FLRW solutions that can be used to model cosmological 
evolution driven by dark energy  (a quintessence field)
and usual matter. In absence of matter, the flow from the 
hyperscaling violating regime to the conformal 
AdS fixed point in  holographic models 
corresponds  to cosmological evolution from  power-law expansion at early 
cosmic times    to  a 
de Sitter (dS)  stable fixed point at late times. In presence of matter, we have a 
scaling regime at early times, followed by an intermediate regime in 
which dark energy tracks matter. At late times the solution exits the 
scaling regime  with a sharp  transition to a  dS spacetime. The phase 
transition between hyperscaling violation and conformal fixed point 
observed in holographic gravity has a cosmological counterpart in  the 
transition between a scaling era  and  a dS era 
dominated by the energy of the vacuum.

\end{abstract}

\author{}
\maketitle
\tableofcontents



\section{Introduction}
Triggered by the anti-de Sitter/Conformal field theory (AdS/CFT) correspondence,  recently we have seen 
several
application of the  
holographic principle aimed  to describe the
strongly coupled regime of quantum field theory (QFT) 
\cite{Hartnoll:2008vx,
Hartnoll:2008kx,
Horowitz:2008bn,Charmousis:2009xr,
Cadoni:2009xm,Goldstein:2009cv,Gouteraux:2012yr}. 
The most interesting example of these applications  is represented by 
the holographic description  of  quantum phase transitions, such as those leading to
critical superconductivity and hyperscaling violation 
\cite{Hartnoll:2008vx,Hartnoll:2008kx,
Charmousis:2009xr,Gubser:2009qt,
Cadoni:2009xm,Goldstein:2009cv,Dong:2012se,Cadoni:2011kv,Huijse:2011ef,
Cadoni:2012uf,Cadoni:2012ea,Narayan:2012hk,Cadoni:2013hna}.

A general question that can be asked in this context is if these 
recent advances can be  used to improve our   understanding, not 
only of some holographic, strongly coupled  dual QFT, but  also  
 of the 
gravitational interaction itself.  After all the holographic 
principle in general and the AdS/CFT correspondence in particular,  
have been often used in this reversed direction. The most important 
example is without doubt the understanding of  the statistical 
entropy of black holes  by counting states in a dual CFT 
\cite{Strominger:1996sh,Strominger:1997eq,Cadoni:1998sg}.

A  challenge for any theory of gravity is surely  cosmology and in 
particular the  
understanding of the present accelerated expansion of the universe 
and the related dark energy hypothesis \cite{Peebles:2002gy,Padmanabhan:2002ji}.  It is  not a priori 
self-evident  that the recent developments on the holographic side 
may be useful for cosmology \cite{McFadden:2009fg}. However,  closer scrutiny reveals that 
key concepts  used in the holographic description can be also used in 
cosmology.

First of all the symmetries of the gravitational 
background. The  AdS and de Sitter (dS)  spacetime in $d$-dimensions share the same 
isometry group (the conformal group in $d-1$ dimensions). This fact 
has been the main motivation  for the formulation of   the dS/CFT 
correspondence \cite{Strominger:2001pn}. Although  this 
correspondence is problematic  \cite{Goheer:2002vf}, it may be very useful 
to relate different gravitational backgrounds if one sees  dS/CFT as 
analytical continuation $r \leftrightarrow it$ of AdS/CFT \cite{Cadoni:2002xe}.

Second,  a domain wall/cosmology 
correspondence  has been proposed \cite{Skenderis:2006jq,Skenderis:2007sm,Shaghoulian:2013qia}. 
For every supersymmetric domain-wall, 
which is solution
of some supergravity (SUGRA) model,  
there is a corresponding flat Friedmann-Lemaitre-Robertson-Walker 
(FLRW) cosmology (which can be obtained  by analytical continuation),
of the same model but with opposite
sign potential.
This means that, although cosmologies in general  cannot be supersymmetric 
they may allow for  the existence of pseudo-Killing
spinors.

Third, the spacelike radial coordinate $r$ of a  static 
asymptotically AdS geometry can 
be interpreted as an energy scale and the corresponding dynamics as a 
renormalization group (RG) flow. This flow drives  the dual QFT from 
an ultraviolet (UV) conformal fixed point (corresponding to the AdS 
geometry) to some nontrivial near-horizon,  infrared (IR)  point where only some 
scaling symmetries are preserved (for instance one can have 
hyperscaling violation in the IR \cite{Cadoni:2013hna}).
By means of the analytic  continuation the RG flow becomes the 
cosmological dynamics  of a time-dependent gravitational background, 
driving the universe from a early time regime  (corresponding to the IR) to 
a late time regime (corresponding to the UV) \cite{Kiritsis:2013gia}.

Last but not least, scalar fields play a crucial role both for holographic 
models and for cosmology. In the first case   they are seen as scalar 
condensates
triggering symmetry breaking and/or phase transitions  in the dual 
QFT \cite{Hartnoll:2008vx,
Hartnoll:2008kx,Cadoni:2009xm}. They are dual to relevant operators that drive the
RG flow from the UV fixed point to the IR critical point. 
Moreover, they  are the sources of scalar solitons, which are the 
gravitational background bridging the asymptotic AdS region and the 
near-horizon region.  On the cosmological side it is well-known that 
scalar fields can be used to model dark energy (the so-called 
quintessence fields) 
\cite{Ford:1987de,Wetterich:1987fm,Caldwell:1997ii,Zlatev:1998tr,amendola}.

In this paper we will consider  a wide class of Einstein-scalar 
gravity model (parametrized by a potential $V$) that have scalar solitonic  
solution interpolating 
between an  hyperscaling violating region and an AdS region. These 
models have been investigated   for holographic applications  
\cite{Charmousis:2009xr,
Cadoni:2009xm,Goldstein:2009cv,Dong:2012se,Cadoni:2011kv,
Cadoni:2012uf,Cadoni:2012ea,Narayan:2012hk,Cadoni:2013hna}.
We show that an analytical continuation transforms  the solitonic 
solution in a flat FLRW solution of a model with opposite sign of $V$.
If the soliton has the AdS region in the UV (IR), the FLRW solution  will  
have a dS epoch at late (early) times. Correspondingly, the FLRW 
solution will   
be characterized by  power-law expansion at early (late) times ( 
Section \ref{sect1}).

Focusing on a particular   Einstein-scalar model (parametrized by a 
 parameter $\beta$) that has  the AdS regime in the UV
and for which exact solitonic solutions are known \cite{Cadoni:2011nq}, we 
generate (and characterize in detail)  the corresponding flat FLRW 
exact solutions. For a broad range of  $\beta$ 
the solutions describe a flat universe decelerating at early times 
but accelerating at late times (Section \ref{sect2}).

We proceed by showing that these solutions can be used  as a model for 
dark energy,  the scalar field playing the role of a quintessence 
field. The parameter of state describing dark energy decreases with 
cosmic time, from  a positive  value ($<1$) till $-1$  (Section \ref{sect3}).

Finally, we  discuss the cosmological dynamics in  presence of matter 
in the form of a general perfect fluid. Although we are not able to 
solve exactly the coupled system, we give strong evidence that the 
universe naturally evolves from a  scaling era at early times  
to a, cosmological constant dominated, 
de Sitter universe  at late times. Moreover, the transition between  
the two regimes in not 
smooth and is the cosmological analogue of the hyperscaling 
violation/AdS spacetime phase transition of holographic models 
\cite{Cadoni:2009xm,Cadoni:2013hna,Gouteraux:2012yr} (Section \ref{sect4}).

\section {Dark energy,  
holographic theories and hyperscaling violation}
\lb{sect1}

We consider Einstein gravity coupled to a real scalar field $\phi$
in four dimensions: 
\begin{equation}
I=\int d^{4}x\sqrt{-g}\left[ {\cal{R}}-\frac{1}{2}\left( \partial \phi \right) ^{2}-V(\phi )
\right] ,  \label{action}
\end{equation}%
 where ${\cal{R}}$ is the scalar curvature of the spacetime.
 The model is parametrized by the self-interaction potential  $V(\phi )$ for the
scalar field.

For static, radially symmetric solutions with planar topology 
for the
transverse space, one can use the following parametrization of the 
solution: 
\begin{equation}\label{metric}
ds^{2}=-U(r)dt^{2}+\frac{dr^{2}}{U(r)}+R^{2}(r)(dx^{2}+dy^{2}),\quad 
\phi=\phi(r).
\eeq
It is known that the theory 
(\ref{action}) admits solutions (\ref{metric}) describing 
black branes   
with scalar hair, at least for 
specific  choices of $V(\phi)$ 
\cite{Cadoni:2011nq,Cadoni:2011yj,Cadoni:2012uf,Cadoni:2013hna}. 
When the spacetime is asymptotically AdS 
\beq\lb{ads}
U=R^{2}= \left(\frac{r}{R_{0}}\right)^{2}
\eeq
(where $R_{0}$ is the AdS length)
or  more generically scale-covariant 
\beq\lb{sc}
U=R^{2}=\left(\frac{r}{r_{-}}\right)^{\eta},
\eeq
(where $r_{-}$ and  $0\le \eta\le 2$ are parameters),  usual no-hair  theorems can be 
circumvented  and regular, hairy black brane solutions of 
(\ref{action}) are allowed \cite{Cadoni:2011nq,Cadoni:2011yj}. 

Moreover, it has been shown that the zero-temperature extremal limit 
of these black brane solutions is necessarily characterized by  $U=R^{2}$ 
in Eq. (\ref{metric}) \cite{Cadoni:2011nq,Cadoni:2013hna}. 
The  extremal limit describes a regular scalar soliton 
interpolating between an AdS spacetime and a scale-covariant metric.
In particular, the  behaviour 
of the potential at $r=\infty$ and in the near-horizon 
region determines the  corresponding geometry. When the leading term of the potential
is a constant $V(\phi)\sim  -6/R_{0}^{2}$ the geometry is AdS. On the 
other hand if the potential behaves exponentially $V(\phi)\sim - 
e^{\lambda \phi}$ ($\lambda$ is some constant) we get a 
scale-covariant metric \cite{Cadoni:2011nq}.

The AdS vacuum has isometries generated by the  conformal group in 
three dimensions. 
In particular  the AdS metric is 
invariant under scale transformations:
\beq\lb{k1}
r\to \mu^{-1} r ,\quad (t,x,y)\to 
\mu(t,x,y).
\eeq

On the other hand the scale-covariant metric breaks some of the 
symmetries of the AdS metric. Under scale transformation the metric 
(\ref{sc}) is not invariant but only scale-covariant. For $\eta\neq 
1$ we get
\beq\lb{k2}
r\to \mu^{\frac{1}{1-\eta}} r ,\quad (t,x,y)\to 
\mu (t,x,y), \quad ds^{2} \to \mu^{\frac{2-\eta}{1-\eta}}ds^{2}.
\eeq

Depending on the form of the potential $V(\phi)$ we have two cases\\
$1)$ AdS is  the  $r=\infty$ asymptotic geometry   and the scale-covariant metric 
is obtained in the near-horizon region \cite{Cadoni:2011nq,Cadoni:2013hna}.\\
$2)$ The AdS spacetime appears in the near-horizon region whereas the scale-covariant metric 
is obtained  as $r=\infty$ asymptotic  geometry \cite{Cadoni:2011yj,Cadoni:2012uf}.\\

This behaviour has a nice holographic interpretation and   a wide range
of application  for describing dual  strongly-coupled 
QFTs and quantum phase  transitions  
 \cite{Hartnoll:2008vx,Hartnoll:2008kx,
Charmousis:2009xr,Gubser:2009qt,Cadoni:2009xm,Cadoni:2011kv,
Cadoni:2012uf,Cadoni:2012ea,Cadoni:2013hna}.

In the dual  QFT the two cases 
described on points $1)$ and $2)$ above correspond, respectively, to the following:\\
$1)$ The dual QFT at zero temperature   has an UV 
conformal fixed point. In the IR it flows to an hyperscaling violating phase 
where the conformal symmetry is broken, only the symmetry  (\ref{k2}) 
is preserved and an IR  mass-scale (the parameter $r_{-}$ in 
Eq.(\ref{sc})) is generated 
\cite{Cadoni:2009xm,Cadoni:2011nq,Cadoni:2011kv,Cadoni:2013hna}.\\
$2)$ The dual  QFT at zero temperature has a conformal fixed point in the 
IR  and flows in the  UV to an hyperscaling violating phase 
\cite{Cadoni:2011yj,Cadoni:2012uf,Cadoni:2012ea}.

When $U=R^{2}$ in Eq. (\ref{metric}) the field equations stemming 
from the action (\ref{action}) become:
\beq\lb{fes}
\frac{R''}{R}=-\frac{\phi'^2}{4},\,\quad\quad 
\frac{d}{dr}(R^{4} \phi')=R^2 \frac{dV}{d\phi},\,\quad\quad
(R^{4})''=-2R^2V(\phi),
\eeq
where the prime denotes derivation with respect  to $r$.
Notice that only two of these equations are independent.

In this paper we are interested in FLRW cosmological solutions with non 
trivial scalar field of the 
gravity theory (\ref{action}). Such solutions have been widely used 
to describe the history of our universe. Depending on the model under 
consideration, the scalar field can be used 
to describe dark energy (quintessence models)\cite{Ford:1987de,Wetterich:1987fm,
Caldwell:1997ii,Zlatev:1998tr,amendola}, the inflaton 
(inflationary models) and also dark matter \cite{Sahni:1999qe,Bertolami:2012xn}.

Our main idea is to use the knowledge of
effective holographic theories of gravity  in the 
cosmological context. The key point is that once an exact static solution 
(\ref{metric}) with $U=R^{2}$  of the field equations (\ref{fes}) is 
known one can immediately  generate a flat FLRW cosmological solution using the 
following transformation in (\ref{metric}) and (\ref{fes}),
\beq\lb{tra}
r\to it,\quad t\to ir,\quad V(\phi)\to -V(\phi).
\eeq
In fact this transformation maps the line element and the scalar 
field (\ref{metric}) into
\begin{equation}
ds^{2}=-R^{-2}(t) dt^{2}+R^2(t)(dr^{2}+dx^{2}+dy^{2}),\quad \phi=\phi(t)
\label{metric1}.
\eeq  
describing   a FLRW metric in which the curvature of 
the  spatial sections is zero, i.e  a flat universe with $R(t)$ playing 
the role of the scale factor.
The same transformation (\ref{tra}) maps the field 
equations (\ref{fes})
into
\beq\lb{fec}
\frac{\ddot{R}}{R}=-\frac{\dot{\phi}^2}{4},\,\quad\quad 
\frac{d}{dt}(R^{4}\dot{\phi})=-R^2 \frac{dV_{c}}{d\phi},\,\quad\quad
\ddot{(R^{4})}=2R^2V_{c}(\phi),
\eeq
where the dot means derivation with respect to the time $t$ and 
$V_{c}=-V$.
One can easily see that Eqs. (\ref{fes}) and (\ref{fec}) have exactly 
the same form, simply with the prime replaced by the dot.
This means that once a zero-temperature static solution, describing a 
scalar soliton, of the theory (\ref{action}) with potential $V$ is 
known, one can immediately write down a cosmological solution of the 
theory  (\ref{action}) with potential $V_{c}=-V$.

The flip of the sign of the potential when passing from the static 
scalar soliton to the cosmological solution has important 
consequences. The AdS vacuum corresponding to constant negative 
potential
$V=-6/R_{0}^{2}$ 
(a negative  cosmological constant) will be mapped 
in the de Sitter   spacetime, corresponding to $V_{c}=6/R_{0}^{2}$ ( 
(a positive cosmological constant), which 
describes an exponentially expanding universe.
Correspondingly, the scale covariant static  metric (\ref{sc}) will 
be mapped into a cosmological power-law solution $R\sim t^{\eta}$.

It follows immediately that the scalar solitons corresponding 
to the cases $1)$ and $2)$ above  will generate after 
the transformation (\ref{tra}) FLRW cosmological solutions with, 
respectively, the 
following properties:

$1)$ The cosmological solution describes an universe evolving from a 
power-law scaling solution at early times  to a  de Sitter spacetime at 
late times.

$2)$ The cosmological solution describes an universe evolving from a 
 de Sitter spacetime
at early times  to a  power-law  solution  at 
late times.

It is interesting to notice that a  universe evolving from a power-law 
solution at early times to an exponentially expanding phase at late 
times has  an holographic counterpart in a QFT flowing from 
hyperscaling violation in the IR to an UV fixed point.
Conversely,  universe evolving from de Sitter at early times to the 
power-law behaviour al late times corresponds to a QFT flowing from  an 
IR fixed point to hyperscaling violation in the UV.

The FLRW solutions described in point $1)$ above are good 
candidates to model an universe, which is dominated at late times by 
dark energy. On the other hand, the cosmological solutions described 
in point $2)$ above are very promising to describe inflation.
In this paper we will investigate in detail solutions of type $1)$.
We will leave the investigation of solution of type $2)$ to a 
successive publication.

Transformations like (\ref{tra}) mapping solitons into FLRW 
cosmologies
 have been already considered in the context of 
SUGRA theories.
\cite{Skenderis:2006jq,Skenderis:2007sm,Shaghoulian:2013qia}. 
They are known under the name of domain wall (DW)/cosmology 
correspondence. 
For every supersymmetric domain-wall, 
which is solution
of some SUGRA model, we can obtain,  by analytical continuation,  
a flat FLRW  cosmology,
of the same model but with opposite
sign potential \cite{Skenderis:2006jq}.

When the model (\ref{action}) is the truncation to 
the metric and scalar sector of some supergravity theory (or more 
generally when the potential $V$ can be derived from a 
superpotential, i.e when  we are dealing with a ``fake'' SUGRA model \cite{DeWolfe:1999cp}) the 
transformation (\ref{tra}) describes exactly the DW/cosmology 
correspondence. However, in this paper we consider the transformation 
in the same spirit of effective holographic theories. We do not 
require the action (\ref{action}) to come from a SUGRA model and we 
consider the transformation (\ref{tra}) in its most general form as a 
mapping between a generic scalar DW solution, i.e.  a spacetime 
(\ref{metric}) with $U=R^{2}$ and cosmological solution 
(\ref{metric1}) endowed with a non trivial time-dependent scalar field.

The cosmological solution (\ref{metric1}) is not written in terms of 
the usual cosmic time $\tau$. Using this time variable, solution (\ref{metric1}) 
 takes the form:
\begin{equation}
ds^{2}=-d\t^{2}+R^2(\tau)(dr^{2}+dx^{2}+dy^{2}),\quad \phi=\phi(\tau)
\label{metric2},
\eeq  
and the coordinate time $t$ and cosmic time $\tau$ are related by
\begin{equation}\lb{tt}
\tau=\int{\frac{dt}{R(t)}}.
\end{equation}

Written in terms of $\tau$ the field equations (\ref{fec}) become the 
usual ones
\beq\lb{fec1}
{\dot H}=-\frac{\dot{\phi}^2}{4},\,\quad 
\ddot{\phi}+3H\dot{\phi}= -\frac{dV_{c}}{d\phi},\,\quad
3H^{2}= \frac{\dot{\phi}^2}{4} + 
\frac{V_{c}}{2}, 
\eeq
where now the dot means derivation with respect to the cosmic time 
$\tau$ and $H$ is the Hubble parameter $H=\dot{R}/R$.
\section{Exact cosmological solutions }
\lb{sect2}
In the previous section we have described a general method that 
allows us to write down a flat FLRW solution with a nontrivial scalar field once a 
static scalar solitonic solution is known. 

In the recent 
literature dealing with  holographic applications of gravity one can 
find several  scalar solitons  describing 
the flow from an scale-covariant metric in the IR to an AdS solution 
in the UV \cite{Cadoni:2011nq,Cadoni:2013hna}.
However, many of them are numeric solutions. An interesting class of 
exact analytic solutions with the above features have been derived in 
Ref. \cite{Cadoni:2011nq} using a generating method. This generating method 
essentially consists in fixing the form of the scalar field. The 
metric part of the solution and the potential $V$ are found by 
solving a Riccati equation and a first order linear equation. 
This allows us to find a solution (\ref{metric}) of the theory 
(\ref{action}) with potential \cite{Cadoni:2011nq}
\begin{equation}\lb{pot}
V_{c}(\phi)=\frac{2}{R_{0}^{2}}e^{2\gamma \beta \phi} \left[2-8 
\beta^2 +(1+8\beta^2)\cosh(\gamma \phi) -6\beta \sinh (\gamma \phi) \right]
\end{equation} 
where \footnote{In 
this paper we are using a normalization of the kinetic term for the 
scalar, which differs from that used in Ref. \cite{Cadoni:2011nq} by a factor of 
$4$. Correspondingly, $\gamma$ differs by a factor of $2$.} 
\beq\lb{para}
\frac{1}{2}\le |\beta|,\quad  \gamma^{-2}=1-4\beta^{2}.
\eeq
The point $\phi=0$ is a maximum of the potential $V$, i.e we have 
$V'(0)=0$ and $V''(0)=-2/R_{0}^{2}=m^{2}<0$, where $m$ is  the 
mass of the scalar field. Notice that the squared-mass of the scalar 
is negative and depends only on the the value of the cosmological 
constant.

The potential (\ref{pot})  contains as  special cases, models 
resulting from truncation to the abelian
sector of $N = 8$, $D = 4$ gauged supergravity \cite{Cadoni:2011nq}. 
In fact, for $\beta=0$ and $\beta=\pm 1/4$ Eq. (\ref{pot}) becomes 
\begin{equation}\lb{sm}
V_{c}(\phi,\beta=0)=\frac{2}{R_{0}^{2}} \left( 2+ \cosh\phi \right), \quad 
V_{c}(\phi,\beta=\pm 1/4)=\frac{6}{R_{0}^{2}}\cosh \left(  
\frac{\phi}{\sqrt{3}}\right).
\end{equation} 
The static, solitonic solutions (\ref{metric}) of the theory (\ref{action}) with 
potential (\ref{pot}) are given by \cite{Cadoni:2011nq}
\begin{equation}\lb{sol}
\gamma \phi=\log X,  \quad R=\frac{r}{R_{0}}X^{\beta 
+\frac{1}{2}},\quad
X=1-\frac{r_{-}}{r},
\end{equation}
where $r_{-}$ is an integration constant.
In the $r=\infty$ asymptotic region, corresponding to $\phi=0$, 
the potential approaches to $-6/R_{0}^{2}$ and 
solutions becomes the AdS solution (\ref{ads}). In the 
near-horizon region, $r=r_{-}$, corresponding to $\phi=\pm \infty$ 
(depending on the sign of $\gamma$), the potential behaves 
exponentially and  the metric becomes, after translation of the $r$ 
coordinate,  the scale 
covariant solution (\ref{sc}) with $\eta=2\b+1$.

A FLRW  solution can be now obtained  applying the transformation 
(\ref{tra}) to Eqs. (\ref{sol}). We simply get 
\begin{equation}\lb{cos}
R(t)=\frac{t}{R_{0}}\left( 1- 
\frac{t_{-}}{t}\right)^{\beta+\frac{1}{2}},\quad \gamma \phi=\log 
\left(1-\frac{t_{-}}{t} \right).
\end{equation}

Solutions (\ref{cos}) is not defined for every real $t$. Moreover, the range of variation of $t$ is 
disconnected. For $t_{-}>0$  we have either  $-\infty <t\le 0$ 
(corresponding to $\g\phi>0$) or $ t_{-}\le t<\infty$ (corresponding to 
$\g\phi<0$). Conversely, for $t_{-}<0$  we have either  $-\infty <t\le t_{-}$ 
(corresponding to $\g\phi<0$) or $ 0\le t<\infty$ (corresponding to 
$\g\phi>0$).

Apart from the parameter $R_{0}$, which sets the value of the 
cosmological constant, the solution (\ref{cos}) depends on the 
parameters $\beta$ and $t_{-}$. The parameter $\g$ is not an 
independent parameter but, apart from the sign,  it is determined  by 
Eq. (\ref{para}).

The potential (\ref{pot}), hence the action (\ref{action}), is invariant under the  
two groups of discrete 
transformations $(\phi\to-\phi, \, \g\to-\g)$ and $(\g\to-\g,\, \b\to 
-\b)$. This symmetries allow to restrict the range of variations of 
$\g,\b$  to  $\{\g>0,\, \phi<0,\, -\frac{1}{2} \le\b\le\frac{1}{2}\}$. 

In terms of the time coordinate $t$ we are left with  only two branches :  $a)\,\{t_{-}>0,\, 
t_{-}\le t<\infty\}$ and $b)\, \{t_{-}<0,\, 
-\infty\le t<t_{-}\}$. However, one can easily realize that these two 
branches are related by the time reversal symmetry $t\to-t,\,t_{-}\to 
-t_{-}$ and are therefore physically equivalent. We are therefore 
allowed to restrict our consideration to the branch $a)$.

The potential $V_{c}$ has a minimum at $\phi=0$. Near the minimum the 
potential behaves quadratically
\beq\lb{qp}
V_{c}= \frac{6}{R_{0}^{2}} + \frac{1}{2}m^{2}\phi^{2}.
\eeq
The squared mass of the 
scalar field is therefore positive and depends only on 
the cosmological  constant 
\beq\lb{sml}
m^{2}= \frac{2}{R_{0}^{2}}= \frac{\Lambda}{3}.
\eeq
As expected, for $t=\infty$ ($\phi=0$)  $V_{c}$ approaches to a positive 
cosmological constant $V_{c}=\Lambda=6/R_{0}^{2}$ and the solution becomes 
the  de Sitter   spacetime. For $t\approx t_{-}$ ( $\phi\to\pm 
\infty$)  the scale factor has a 
power-law form, $R\propto (t-t_{-})^{\beta+1/2}$ and the potential 
behaves exponentially. We get, respectively for $\phi=\pm \infty$, the 
asymptotic behaviour 
\bea\lb{ab}
V_{c}(\phi)&=&  R_{0}^{-2}(1+8\beta^2-6\beta)e^{\gamma \phi 
(2\beta+1)},\nonumber\\   
V_{c}(\phi)&=& R_{0}^{-2}(1+8\beta^2+6\beta)e^{\gamma \phi 
(2\beta-1)}.  
\eea

The range of variation of the parameter $\beta$ can be  further constrained 
by some physical requirements that 
must be fulfilled if solution (\ref{cos}) has to describe the late-time 
acceleration  of 
our universe. 

The usual way to achieve this is  to   considers quintessence  models  
characterized by a slow roll of the scalar field. As we will see 
later in this paper  the 
potential (\ref{pot}) does not satisfy the
slow roll conditions, which are sufficient,  but not necessary, 
for having late-time  acceleration. 
We will use here a much weaker condition on the slope of the 
potential $V_{c}(\phi)$.

The scalar field $\phi$ in Eq. (\ref{cos}) 
is a monotonic function of the time $t$ in the   branch under 
consideration.  Being the function $\phi(t)$  of Eq. (\ref{cos}) 
monotonic for $t_{-}>0$ 
and $t\in (t_{-}, \infty)$  the simplest way to have a well-defined 
physical model (i.e a one-to-one correspondence $t \leftrightarrow 
V_{c}$)   is to require also the potential to be a  
 monotonic function  inside the  branch.
This requirement restricts the range of variation
 of the parameter $\beta$ to

\beq\lb{rv}
-\frac{1}{4}<\b\le\frac{1}{4}.
\eeq
In fact, for $\frac{1}{4}<|\b|\le\frac{1}{2}$ 
the potential $V_{c}$ has  other extrema. From the range of $\beta$, we 
have excluded the point 
$\beta=-1/4$ because in this case the potential (\ref{pot}) becomes  
exactly the same as for $\beta=1/4$.  It is interesting to notice that 
the two simple models (\ref{sm}), arising from SUGRA truncations, 
appear as the two limiting cases of this range of variation.

In conclusion, the  FLRW solution (\ref{cos}) represents a well-behaved cosmological solution in the  
following range of the 
parameters and of the time coordinate $t$
\beq\lb{pr}
-\frac{1}{4}< \b\le \frac{1}{4},\quad 1\le \g\le \frac{2}{\sqrt 
3},\quad t_{-}>0,\quad
t_{-}\le t<\infty,\quad \phi<0.
\eeq
Other branches are either physically equivalent  to it (by using the 
discrete symmetries of the potential (\ref{pot}) or time-reversal 
transformations) or can be excluded by physical arguments.

Let us now consider the Hubble 
parameter $H$ and the acceleration parameter $A$. 
We have for $H$  and  $A$:
\bea\lb{ha}
H&=&\frac{1}{R}\frac{dR}{d\tau}=\frac{dR}{dt}=
\frac{X^{\alpha}}{R_{0}}\left[1+
\alpha X^{-1}\left(\frac{t_{-}}{t}\right) \right],
\nonumber\\
A&=&\frac{1}{R} 
\frac{d^{2}R}{d\tau^2}=\left(\frac{dR}{dt}\right)^{2}+R\frac{d^{2}R}{dt^2}=
\frac{X^{2\alpha -2}}{R_{0}^{2}t^{2}} 
\left[\left( t+ (\alpha-1)t_{-}\right)^{2} + \alpha(\alpha-1)t_{-}^{2}\right].
\eea
where $\a=\b +\frac{1}{2}$. 
An important physical requirements are the positivity of the Hubble 
parameter $H$. Moreover, the acceleration parameter $A$ must be 
positive, at least at late times, to describe late-time  
acceleration.

One can easily check that in the range of variation of 
the parameter $\b$ (\ref{pr})  we 
have always $H>0$. The behaviour of the acceleration parameter $A$ is 
more involved.  $A$ becomes zero for $t_{12}= 
\left[1-\a\pm\sqrt{\a(1-\a)}\right]t_{-}$. For $t_{-}>0$ we have 
$t_{1}> t_{-}, \, t_{2}<t_{-}$ for $-1/4<\b\le 0 $, whereas  $t_{1}< 
t_{-}, \, t_{2}<0$ for $0\le\b\le  
1/4$.  This means that in the branch under consideration for $\b$ positive, 
the universe is always 
accelerating. For $\b$ negative  the universe will have  a deceleration at 
early times (for $t_{-}<t<t_{1}$), whereas it will accelerate for 
$t>t_{1}$.

Until now we have always used in our discussion the coordinate time 
$t$. The cosmic time $\tau$ is defined implicitly in terms of $t$ by Eq. (\ref{tt}). 
The correspondence $\tau\leftrightarrow t$ defined by Eq. (\ref{tt}) must be 
one-to-one, i.e $\tau(t)$ must be monotonic in the range (\ref{pr}).
Let us show that this is indeed the case.
Inserting  the expression for $R$ given in  Eq. (\ref{cos}) into 
(\ref{tt}) we get
\beq\lb{tint}
\frac{\tau}{R_{0}}= \int \frac{dt}{t} 
\left(\frac{t}{t-t_{-}}\right)^{\b+\frac{1}{2}}= -B_{z}(0, 
\frac{3}{2}-\b),
\eeq
where $B_{z}(0, 
\frac{3}{2}-\b)$  is the incomplete  beta function $B_{z}(p,q)$ and $z= t_{-}/t$.
From the previous equation we get  the leading behaviour of $\tau(t)$ 
near $t=t_{-}$ and $t=\infty$. We have, respectively,
\beq\lb{lb}
\tau\propto (t-t_{-})^{\frac{1}{2}-\beta}, \quad \tau\propto \log t.
\eeq
From this equation we learn that $t=t_{-}$ and $t=\infty$ are mapped, 
respectively into $\tau=0$ and $\tau=\infty$.  Moreover, from Eq. 
 (\ref{tint}) one easily realises that $d\tau/dt$ is always 
strictly positive for $t_{-}\le t <\infty$.

When $\beta$ is a generic real number  in $(-\frac{1}{4},\frac{1}{4})$ the 
function $\tau(t)$ cannot be expressed in terms of elementary 
functions. However, the integral (\ref{tint}) can be explicitly 
computed when $\b$ is a rational number.
The simplest example is given by $\beta=0$. 
In this case we get for the function $t=t(\tau)$, the scale factor 
$R$ and the scalar field $\phi$,
\beq\lb{f1}
\frac{t}{t_{-}}= \cosh^{2}\frac{\tau}{2R_{0}},\quad 
R(\tau)=\frac{t_{-}}{2R_{0}}\sinh \frac{\tau}{R_{0}},\quad \phi= 
2\log \tanh \left(\frac{\t}{2R_{0}}\right).
\eeq
An other simple example is obtained for $\b=1/4$.
We  get 
\beq\lb{f2}
\frac{\tau}{R_{0}}= -2 \arctan Y-\log 
\frac{Y-1}{Y+1} +\pi,\,\quad
\quad Y^{4}=\frac{t}{t-t_{-}}.
\eeq
Let us conclude this section by giving a short description of the 
evolution of our universe described by Eq. (\ref{cos}).

The universe starts from a curvature singularity at $\tau=0$, where  
the scale factor vanishes, $R=0$, and the scalar field, the Hubble 
parameter and the acceleration  diverge.

For $\tau>0$ the potential $V_{c}(\phi$) rolls down to its minimum at 
$\phi=0$ first following the exponential behavior given by Eq.  
(\ref{ab}).  In this early stage the scale factor evolves following a 
power-law behaviour whereas the scalar field evolves logarithmically:
\beq\lb{sca}
R\sim \tau^{\frac{1+2\b}{1-2\b}},\quad H\sim \frac{1}{\tau},\quad 
A\sim \frac{1}{\tau^{2}},\quad \phi\sim \log \tau.
\eeq
The acceleration $A$ is positive for  $\b>0$ and negative for $\b<0$.
After a time-scale determined by $t_{-}$ the universe  enters, for 
$\b$ negative,
in an accelerating phase, 
whereas for $\b$ positive continues to accelerate.

At late times, independently of the value of $\b$, the potential approaches 
the quadratic minimum at 
$\phi=0$ and the universe has an exponential expansion described by 
de Sitter spacetime and a constant scalar. Therefore at late times  
the universe forgets about its initial conditions (the parameter 
$t_{-}$)  and all the physical 
parameters are determined completely in terms of the cosmological 
constant. We have for 
the mass of the scalar field and for  $H,A$:
\beq\lb{ds1}
m^{2}=2H^{2}=2A=\frac{2}{R_{0}^{2}}=\frac{\Lambda}{3}.
\eeq
This behaviour is the  cosmological counterpart  of the flowing to  
an UV   conformal fixed point of solitonic solutions in effective holographic 
theories with an hyperscaling violating phase. 
The dS solution  corresponds to AdS  vacuum (\ref{ads}) and is invariant under the 
scale symmetries (\ref{k1}) (obviously exchanging the $r,t$ 
coordinates). The power-law solution (\ref{sca}) corresponds to the scale 
covariant solution (\ref{sc}), it shares with it the scale
symmetries (\ref{k2}).

Thus, both class of solutions (the scalar soliton and the 
cosmological solutions)  are characterized by the emergence 
of a mass-scale. In the case of the scalar soliton (\ref{sol}) this 
mass-scale is described by the the parameter $r_{-}$ and emerges in 
the  IR  of the dual QFT. In the case of the cosmological solution 
the mass-scale  is described by the the parameter $t_{-}$, which 
characterizes the early-times cosmology.

When the dual QFT   flows in the UV fixed 
point,  the conformal symmetry washes out all the information about 
the  IR length $r_{-}$ which, characterizes the hyperscaling 
violating phase \cite{Dong:2012se,
Cadoni:2012uf}. Similarly, the cosmological evolution 
washes out all the information about the initial parameter $t_{-}$ and 
all the physical parameters are completely  determined  by the  
cosmological constant.

In the next sections we will show how our cosmological solutions can 
be used to model dark energy.

\section{Dark energy models}
\lb{sect3}
It is well known that dark energy can be considered   a modified form 
of matter. The simplest way   to model it, is  by means of a scalar field 
(usually called quintessence)
coupled to usual Einstein gravity, i.e with  a model given by 
(\ref{action}) with properly chosen potential. 

Modelling dark energy with a scalar field has many advantages. Unlike 
the cosmological constant scenario,  the 
energy density of the scalar field at early times does not necessarily need to be 
small with respect to the other forms of matter. 
Cosmological evolution can be described as a dynamical system. It 
allows for the existence of attractor-like solutions 
(the so called ``trackers'')
in which the energy density of the scalar field is comparable with 
the the usual matter-fluid density for a wide range of initial 
conditions.  This helps to solve  the so-called coincidence problem 
of dark energy (see e.g. \cite{amendola}).

The model described by Eq. (\ref{action}) with the potential 
(\ref{pot}) is  a good candidate for realizing a tracking behaviour.
In fact, at early times the potential behaves exponentially (see 
Eq. (\ref{ab})) giving the power-law cosmological solution (\ref{sca}).
This kind of solution have been widely used to produce  tracking 
behavior at early times. 
Moreover, at late times our model flows in a  dS solution (i.e  a 
solution modelling dark energy as a cosmological constant). This 
could help to explain the present accelerated expansion  of the 
universe characterized  by the tiny energy scale $\Lambda\approx 
10^{-123} m_{pl}^{2}$.

Obviously,  to be realistic our  models must pass all the tests 
coming from cosmological observations. The most stringent coming from 
the  above value of the cosmological constant.  

In this section we will address  the issues sketched above for our 
cosmological model (\ref{pot}).

Being dark energy described  as an exotic form of matter, useful 
information comes from its   equation of state 
$p_{\phi}=w_{\phi}\rho_{\phi}$. For a 
quintessence model described by the action (\ref{action}) one has 
\beq\lb{eqs}
w_{\phi}=\frac{p_{\phi}}{\rho_{\phi}}= 
\frac{T(\phi)-V_{c}(\phi)}{T(\phi)+V_{c}(\phi)}=\frac{1-K(\phi)}{1+K(\phi)} .
\eeq
where $T(\phi)=\dot{\phi}^{2}/2$ (the dot means derivation with 
respect the cosmic time $\tau$) is the kinetic energy of the scalar 
field and we have defined $K(\phi)=V_{c}/T$ as the ratio between 
potential and kinetic energy. 
The expression of  $T$ and $K$ as a function  of  $\phi$ can be 
easily computed using Eq. (\ref{cos}) and (\ref{tt}).  We have 
$\frac{t}{t_{-}}= (1- e^{\g \phi})^{-1}$ and $T(\phi)= 
\frac{2}{(R_{0}\g)^{2}}e^{2\g \b \phi}\sinh^{2}(\g 
\phi/2)$.  Whereas for $K$ we obtain
\beq\lb{ke}
 K(\phi)= \g^{2} \frac{ 2-8\b^{2}+( 1+8 \b^{2}) \cosh\g 
\phi- 6\b\sinh\g 
\phi}{\sinh^{2}\frac{\g 
\phi}{2}}.
\eeq
From these equations one can easily  derive the time evolution of the 
parameter of state $w_{\phi}$.  At $\tau=0$, corresponding to 
$\phi=-\infty$, both the kinetic and potential energy, as a function 
of $\phi$, diverge 
exponentially but their ratio is constant. $w_{\phi}$   takes the 
$\beta$-dependent value
\beq\lb{wo} 
w_{0}(\beta)= -\frac{1+10\b}{3(1+2\b)}.
\eeq 
In the range of  variation of $\b$ we have  $-7/9\le w_{0}< 1$.
In particular, for  $0\le \b \le 1/4$, $w_{0}(\b)$ is always negative  
($-7/9\le w_{0}\le - 1/3$), whereas for $-1/4< \b \le 0$ ,  $w_{0}(\b)$ 
goes from $-1/3$ to $1$.  
For $0<\tau <\infty$ (corresponding to $-\infty <\phi<0$) the ratio 
$K$  increases and,  correspondingly, $w_{\phi}$  decreases, monotonically 
from $w_{\phi}=w_{0}(\b) $ to $w_{\phi}=-1$. At $\tau=\infty$ ($\phi=0$) 
the potential energy goes to a minimum, the kinetic energy 
vanishes  and the state parameter $w_{\phi}$ attains the value 
corresponding to a cosmological constant $w_{\phi}=-1$.

As expected  dark energy has an equation of state with $-1\le 
w_{\phi}< 1 $ negative, 
but bigger than $-1$. The  $w_{\phi}=-1$ value, corresponding to a 
cosmological constant,  is attained when the potential rolls in its 
$\phi=0$ minimum  at $\tau=\infty$.

The  behaviour of the parameter $w_{\phi}(t)$  is perfectly consistent with 
what we found for the acceleration parameter $A$. In fact, for $\b$ 
positive $ -1\le w_{\phi}(t)<-1/3$ and the universe always accelerates.
For $\b$ negative,  $ -1\le w_{\phi}(t)< 1$  and we have a transition from  
early-times deceleration ($w_{\phi}(t)>-1/3$) to late-times acceleration 
($w_{\phi}(t)<-1/3$).

As we have mentioned in the previous section, in our model, late-time 
acceleration is not produced by the usual mechanism used in quintessence 
models, i.e by a slow-roll of the
scalar field. Late-time acceleration requires  $w_{\phi}<-1/3$ hence from 
Eq. (\ref{eqs}), $K=V_{c}/T>2$. Sufficient conditions to satisfy the 
latter inequality is a slow evolution of the scalar field, which is 
guaranteed by the slow-roll conditions \cite{Bassett:2005xm}
\beq
\lb{ep}
\e= \left( \frac{1}{V_{c}}\frac{dV_{c}}{d\phi}\right)^2 \ll 1,\quad 
|\mu|=2 \left| \frac{1}{V_{c}}\frac{d^{2}V_{c}}{d\phi^{2}} \right| \ll 1. 
\eeq
In our model, the potential $V_{c}$ at late times behaves as 
Klein-Gordon potential (\ref{qp}), so that we have:

\beq \lb{em}
\e=\m=\frac{4}{\f^2}
\eeq
Obviously  the slow-roll parameters \ref{em} go to 
infinity at late time when $\phi$
approaches to $0$.  
However, the slow-roll conditions (\ref{ep}) are sufficient but not 
necessary for having late-time  acceleration. In our model the 
condition $V_{c}>2T$ is satisfied by an alternative (freezing) mechanism: at late 
times the scalar field approaches its minimum at $\phi=0$ in which 
the potential energy $V_{c}$ is constant and non-vanishing  whereas  the kinetic energy $T$ 
is zero. 

\section{Coupling to matter}
\lb{sect4}
Until now we have considered a quintessence model (\ref{action}) with 
the potential (\ref{pot}) and shown that for a wide range of  the  
parameter $\b$ it can be consistently  used to produce a late-time 
accelerating universe. The next step is to introduce  matter fields 
in the action, in the form of a general perfect fluid 
(non-relativistic matter or radiation). Obviously, this is a crucial 
step because the key features of quintessence model (tracking 
behavior, stability etc.) are related to the presence of matter.

In presence of matter the cosmological equations can be written as
\beq\lb{fec2}
{\dot H}=-\frac{1}{4}\left(\rho_{\phi}+\rho_{M}+p_{\phi}+p_{M}\right),\,\quad 
\ddot{\phi}+3H\dot{\phi}= -\frac{dV_{c}}{d\phi},\,\quad
H^{2}= \frac{1}{6}\left(\rho_{\phi}+\rho_{M}\right), 
\eeq
where $\rho_{\phi}= {\dot \phi}^{2}/2 +V_{c},\, p_{\phi}= {\dot 
\phi}^{2}/2 -V_{c}$ are the density and pressure of the quintessence 
field, whereas $\rho_{M}$ and $p_{M}$ are those of matter, related by 
the equation of state $p_{M}=w_{M}\rho_{M}$.

The cosmological dynamics following from Eqs. (\ref{fec2}) can be 
 recast in the form of  a dynamical system.  By defining 
$x={\dot\phi}/(\sqrt{12} H),\, y= \sqrt{V_{c}}/(\sqrt{6}H),\, N=\log R$, the 
cosmological equations (\ref{fec2}) take the form (see e.g.  
\cite{amendola}):

\bea\lb{ds}
\frac{dx}{dN}&=& - 3x+ \sqrt{\frac{3}{2}}\l y^{2} + \frac{3}{2} 
x\left[\left(1-w_{M}\right)x^{2}+\left(1+w_{M}\right)\left(1-y^{2}\right)\right]\nonumber\\
\frac{dy}{dN}&=& - \sqrt{\frac{3}{2}}\l xy+  \frac{3}{2} 
y\left[\left(1-w_{M}\right)x^{2}+\left(1+w_{M}\right)\left(1-y^{2}\right)\right]\nonumber\\
\frac{d\l}{dN}&=& - \sqrt{6} \l^{2}\left(\Gamma -1\right) x,\quad \l= 
- \frac{\sqrt{2}}{V_{c} }\frac{dV_{c} }{d\phi},\quad \Gamma =V_{c} 
\frac{d^{2}V_{c} }{d\phi^{2}}\left(\frac{dV_{c} }{d\phi}\right)^{-2}.
\eea
This form of the 
dynamics is particularly useful for investigating the fixed points 
of the dynamics  and their stability.
In the case of a potential given by Eq. (\ref{pot}) neither $\l$ nor 
$\Gamma$ are constant and Eqs. (\ref{ds}) cannot be 
solved analytically. Even the characterization of the fixed points of 
the dynamical system is rather involved. 

To gain information about the cosmological dynamics we will use a 
simplified approach. We will first consider the dynamics in the two 
limiting regimes of small and large cosmic time, i.e  $(1) \,\,\phi 
\to -\infty,\quad (2)\,\, \phi=0$ in which the 
potential behaves, respectively, exponentially (see Eq. (\ref{ab}) 
and quadratically (see Eq. (\ref{qp})) and the scale factor evolves, 
respectively, as power-law and exponentially. After that we will describe 
qualitatively the cosmological evolution   
in the intermediate region $\phi\approx -1/\gamma$.
\subsection{Power-law evolution} 
In the case of an exponential potential $\l=const$ in Eq. (\ref{ds}). 
Both the fixed points of the 
dynamical system (\ref{ds}) and their stability  are well known  
\cite{Copeland:1997et,Neupane:2003cs,amendola}). Apart from  fluid-dominated and 
quintessence-kinetic-energy-dominated fixed points, which are not 
interesting for our purposes, we have two fixed points in  
which the scale factor $R$ has a power-law behavior.

The first fixed point is obtained for
\beq\lb{kk}
x=\frac{\l}{\sqrt{6}},\quad y=\sqrt{1-\frac{\l^{2}}{6}},\quad 
\l=\sqrt{\frac{2(1-2\b)}{1+2\b}},
\eeq
describes a 
quintessence-dominated solution with 
$\Omega_{\phi}=\frac{\rho_{\phi}}{6H^{2}}=1$ and a constant parameter 
of state $w_{\phi}=w_{0}{(\b})$ with $w_{0}{(\b})$  given by Eq. 
(\ref{wo}). This fixed point is stable   for  
\beq\lb{op}
\b>\b_{0}= -\frac{1+3w_{M}}{2(5+3w_{M})}.
\eeq
Notice that if we take matter with $ 0\le w_{M}<1$ we have $ 
-1/4< \b_{0}\le- 1/10$ so  that the region of stability is inside 
the range of definition of the parameter $\beta$. 
One can easily realize that this solution is nothing but the 
previously found power-law solution (\ref{sca}) with  the constant parameter of state 
$w_{0}(\b)$ given by Eq. (\ref{wo}). Because $\Omega_{\phi}=1$ this solution cannot 
be obviously used to realize the radiation or matter-dominated 
epochs.

Phenomenologically more interesting is the second fixed point of 
the dynamical system (\ref{ds})  with an exponential potential.
This is the so-called scaling solution \cite{Copeland:1997et,Liddle:1998xm} and is given by
\beq\lb{kl}
x=\sqrt{\frac{3}{2}}\frac{(1+w_{M})}{\l},\quad 
y=\sqrt{\frac{3(1-w_{M}^{2})}{2\l^{2}}},
\eeq
where $\l$ is given as in Eq. (\ref{kk}).
This scaling solution is characterized by a constant ratio 
$\Omega_{\phi}/\Omega_{M}$  and by the equality of the parameter of 
state for quintessence and matter $w_{\phi}=w_{M}$. Moreover we have
\beq\lb{fg}
\Omega_{\phi}=x^{2}+y^{2}= \frac{3}{\l^{2}}(1+w_{M}).
\eeq
The scale factor $R$ behaves also in this case as a power-law, with a 
$w_{M}$ dependent exponent, $R\propto \tau^{2/(3(1+w_{M}))}$.

The scaling solution is a stable attractor for $ \b_{1}<\b<\b_{0}$, where $\b_{0}$ 
is given as in Eq. (\ref{op}) and  
\beq\lb{rd}
\b_{1}= -\frac{12w_{M}^{2}+15w_{M}+5}{2(12w_{M}^{2}+33w_{M}+19)}.
\eeq
Notice that for ordinary matter  characterized  $0\le w_{M}<1$ we 
have  $ -1/4 <\b_{0}\le -1/10$ and $ -1/4 <\b_{1}\le -5/38$. Hence, the range 
of stability of the scaling solution is well inside the range of 
definition of $\beta$.  For  $\b>\b_{0}$ the scaling solution is a 
saddle point, whereas for $\b<\b_{1}$ it is a stable spiral.

The scaling solution has features that make it very appealing for 
describing the early-time universe. The ratio 
$\Omega_{\phi}/\Omega_{M}$ is constant and $\lambda$-dependent, in 
principle  $\l$ can be 
chosen in such way that $\Omega_{\phi}$ and $\Omega_{M}$ have the 
same order of magnitude. Moreover the solution is an attractor 
making  the dynamics  largely independent of the 
initial conditions. These features allow to solve the coincidence 
problem. Cosmological evolution will be driven   sooner or later
to the scaling fixed   point, allowing to have a value of density of 
the scalar field of 
the same order of
magnitude of matter (or radiation) at the ending of inflation.

Despite these nice features the scaling solution alone cannot be 
used to model the matter-dominated epoch of our universe  for several reasons. 
Because  
$w_{\phi}=w_{M}$ it is not possible to realize cosmic acceleration 
using a scaling solution. The universe must therefore exit the 
scaling  era, characterized by $\Omega_{\phi}=constant$, to connect 
to the accelerated epoch, but this is not possible if the parameters  
are within the range of stability of the solution. An other problem 
comes from nucleosynthesis constraints. They require  
$\Omega_{\phi}/\Omega_{M}<0.2$. However, in the range of the 
parameter $\beta$ where 
the scaling solution is a stable node  the minimum value of the ratio is given by 
$\Omega_{\phi}/\Omega_{M}= (7+9w_{M})/(1-w_{M})$. In the most 
favourable case, $w_{M}=0$ 
(non-relativistic matter), we still have $\Omega_{\phi}=7\Omega_{M}$.
The situation improves if we move in the region where the scaling 
solution is a stable spiral. Taking $w_{M}=0$ 
we find  $1<\Omega_{\phi}/\Omega_{M}<7$, with 
$\Omega_{\phi}/\Omega_{M}\to 1$ for $\b\to -1/4$.

In the model under consideration some of these difficulties have the chance to 
be solved because the dynamics exits naturally the scaling era, at 
times when the exponential approximation $\l \approx const.$ 
 is not 
anymore valid. 

\subsection{Exponential evolution}
\lb{ssect1}

At late cosmic times the scalar field potential  behaves 
as in Eq. (\ref{qp}) and the dynamics  of the scalar field  
is governed  by the equation:
\beq \lb{ev}
\ddot{\phi}+3H\dot{\phi}+m^{2}\phi=0,
\eeq
which is can be considered as describing a damped harmonic oscillator. In this  analogy 
the scalar 
mass $m$ represents the pulsation of the oscillations and the Hubble parameter $H$
acts as a (Hubble) friction term. Two cases are possible 
\cite{Turner:1983he,Dutta:2008px}: 

$(a)$ 
$r=3H/m>1$, the oscillations are suppressed 
by  Hubble friction   and  $\f$ goes to a constant value (overdamping);

$(b)\,$ 
$r \ll 1$,  the oscillating  term dominates over Hubble friction  
and the scalar field  oscillates around the minimum of the potential.

Depending  on the global dynamics of the system either  case $(a)$ or 
case $(b)$ will be realized.  Presently  we do not have an 
exact  control of this global dynamic. By studying the intermediate 
regime, however,   we will give in Sect.
\ref{ssect2},  strong evidence that cosmological evolution will be 
driven near to  de Sitter point where  $\phi\approx 0$
 
In the limit $\phi\to 0$ we have $V_{c}/\dot \phi^{2}\gg1$ and the scalar 
field is frozen to a constant value and one can easily see that case 
$(a)$ is realized.
This can be also checked  directly. From Eq. (\ref{ds1}) 
 we can  easily read out the ratio 
$r=\frac{3}{\sqrt{2}} >1$ for our mode model, so that we have 
overdamping.

The absolute value of the scalar field
decreases and approaches asymptotically the minimum of the 
potential where we can approximate $V_{c}(\phi)$ by a constant.
Moreover, the value of the  ratio $r$ does not depend  on the  
parameter $\beta$. The value of the scalar field is completely 
determined by Eq. (\ref{ev}) and, in particular, is independent from 
the early time dynamics. This is again a manifestation of the 
conformal and scaling symmetries of the gravitational background: 
once the cosmological dynamics is driven near to the  de Sitter vacuum any 
memory about the scaling regime is lost, the dynamics becomes 
universal and depends only on one mass-scale, that is set by the 
cosmological constant. 

This behaviour has to be compared with that pertinent to the 
previously discussed  slow-roll conditions (\ref{ep}).  
They correspond to have $V_{c}  \gg \dot{\f^2}$ and $|\ddot{\f}| \ll 
|3H\dot{\f}|$ in (\ref{ev}). We can produce in this way late-time 
acceleration but the late-time dynamics is not universal but depends 
on the details of the model. 

Because of overdamping the cosmological evolution will be driven near 
to the minimum of the potential $V_{c} (\phi)$. In this region the 
potential at leading order can be approximated by a cosmological 
constant, 
$V_{c} (\phi)= 6/R_{0}^{2}$.
For a constant potential we have $\l=0$ in Eq. (\ref{ds}) and we can 
easily find the fixed points of the dynamical system. 

We have three 
fixed points $(1)$  $x=y=\Omega_{\phi}=w_{\phi}=0,\, \Omega_{M}=1,$
which represents  a fluid-dominated solution. $(2)$  $y=0,\, x=\pm 
1,\, \Omega_{\phi}= w_{\phi}=1,\, \Omega_{M}=0,$ which represents a 
 solution dominated by the  kinetic energy of the scalar field. $(3)$  $x=0,\, y=\pm 
1,\, \Omega_{\phi}=1,\, w_{\phi}=-1,\, \Omega_{M}=0,$ which represents a 
a solution dominated by the energy of the vacuum (cosmological 
constant). Obviously, the only physical candidate for describing the 
late-time evolution of our universe is fixed point $(3)$. 

Neglecting 
the solution with negative $y$ (representing an exponentially 
shrinking universe), the solution with $y=1$ give the de Sitter 
spacetime,  an exponentially expanding universe with 
$H=R_{0}^{-1}$, i.e. $R\propto \exp{\tau/R_{0}}$.
By linearizing Eqs. (\ref{ds}) around the fixed point, one can easily find that the de Sitter 
solution is a stable node  of the dynamical system. In fact the two  
eigenvalues of  the matrix describing the linearized system are  real 
and negative ($-3, -3(1+w_{M})$).

Actually, for $\l=0$ one can go further and integrate exactly  the 
dynamical system (\ref{ds}). After some calculation  one finds 
\beq\lb{sol4}
y=\frac{1}{\sqrt{1+c R^{-3(1+w_{M})} -a^{2}R^{-6}}},\quad
x=\frac{a R^{-3}}{\sqrt{1+c R^{-3(1+w_{M})} -a^{2}R^{-6}}},
\eeq
where $a,c$ are integration constants. 

Eq. (\ref{sol4}) confirms that 
the dS spacetime is an attractor of the dynamical system. In fact, 
the two-parameter family of solutions (\ref{sol4}) has a node at 
$x=0,y=1$ to which every member of the family approaches as $R\to \infty$. 
The three terms in the square 
root in the denominator represent, respectively,  the contribution of the energy of the 
vacuum, the contribution of matter, and the 
contribution of the kinetic energy of the scalar field.
One can easily see that at late times ($R\to \infty$) the vacuum energy 
always dominates over the other two contributions. Moreover, the 
scalar field kinetic energy contribution is always subdominant with 
respect to the 
matter contribution. In absence of matter ($c=0$) we have  
$HR_{0}=\sqrt{1- a^{2} R^{-6}},\quad \dot\phi^{2}\propto R^{-6}$, 
telling us that the kinetic energy of the scalar field falls off very 
rapidly as the scale factor $R$ increases.

An explicit form of the time dependence of the  scale factor  can be 
derived from (\ref{sol4})  only after fixing the parameter of state 
of matter. For dust ($w_{M}=0$) and  radiation ($w_{M}=1/3$) we find, 

\beq\lb{sol9}
R_{dust}(\tau)= c_{1}\left[\cosh \frac{3}{2R^{0}}(\tau-\tau_{0})\right]^{\frac{2}{3}}, 
\quad R_{rad}(\tau)= c_{2}\left[\cosh 
\frac{2}{R^{0}}(\tau-\tau_{0})\right]^{\frac{1}{2}},
\eeq
where $c_{1,2}, \tau_{0}$ are  constants.

Summarizing, if cosmological evolution is such that the system is 
driven near to the minimum of the potential $V_{c}$, i.e the  region where the potential 
can be approximated by a cosmological constant, then the universe will 
necessarily enter in the regime of exponential expansion described 
by the dS spacetime.
Obviously, the crucial question is: will the system be driven to this 
near-minimum region?  A definite answer to this question requires a 
full control of the global dynamics of the system (\ref{ds}). 
In the next subsection, by analyzing the intermediate region of the 
potential $V_{c}$,  we will give strong indications that this is 
indeed the case.

\subsection{Intermediate regime}
\lb{ssect2}

A key role in discussing cosmological evolution in presence of dark 
energy is played by the so-called tracker solutions \cite{Steinhardt:1999nw}.
These solutions are special attractor trajectories in the phase space of the 
dynamical system (\ref{ds}) characterized by having approximately 
constant $\l,\Omega_{\phi},w_{\phi}$. If the time-scale of the 
variation of $\l$ is much less then $H^{-1}$ we can consider these 
trajectories as build up from instantaneous fixed points changing in 
time \cite{Steinhardt:1999nw,amendola}.  Thus, tracker solutions  are very useful to solve the coincidence 
problem. During the matter dominate epoch dark energy tracks matter, the ratio  
$\Omega_{\phi}/\Omega_{M}$  remains almost 
constant and $w_{\phi}$ remains close to $w_{M}$ with 
$w_{\phi}<w_{M}$. 
Moreover, if the condition $\Gamma>1$ along the 
trajectory is satisfied, $\l$ decreases toward zero. Once the the 
value $\l^{2}= 3(1+w_{M)}$ is reached the  fixed point (\ref{kk}) with 
$\Omega_{\phi}=-1$ becomes stable and the universe exits the scaling 
phase to enter the accelerated phase. 

To check if our solutions behave as  trackers let us first calculate 
the parameter $\Gamma$ of Eq. (\ref{ds}) for our potential 
(\ref{pot}). We get
\beq\lb{gm1}
\Gamma -1= \g^{2}\frac{1-16\b^{2}+2 (1+8\b^{2})\cosh \g \phi-12 \b 
\sinh \g \phi}{\left(4\b-4\b \cosh\g \phi+\sinh\g\phi\right)^{2}}.
\eeq
One can check analytically and numerically that  for $ -1/4 <\b\le 
1/4$  we have $\Gamma-1=0$ for $\phi=-\infty$.
In the range $\phi\in (-\infty 
,0)$, $\Gamma-1$  monotonically increases and blows up to $\infty$ as 
$1/\phi^{2}$ for 
$\phi=0$. 
In Figs. \ref{fig:lG} we show the plot of  $\l$ and $\Gamma-1$ as 
a function of $\phi$ for selected values of the parameter $\b$. The 
curves remain flat till the scalar field reaches values of order 
$-1/\g$. 

Moreover $\Gamma-1$  is exponentially suppressed as 
$\phi\to-\infty$ and stays flat, near to zero,  till we reach values 
of $|\phi|$ of order $1/\g$. For instance for  $-\infty <\phi<-10/\g$ 
we have $0<(\Gamma-1)<10^{-4}$. This shows that  in the range 
$(-\infty, {\cal{O}}(1/\g))$,  $\Gamma$ varies 
very slowly as a function of $\phi$. The same is true if we consider 
$\Gamma$ as a function of the number of e-foldings $N$. In fact we 
have  $d\Gamma/dN= \sqrt{12}x(d\Gamma/d\phi)$ and because $x$ flows 
from the value $x= \sqrt{3/2} (1+w_{M})/\l$ at the scaling fixed 
point to $x=0$ at the dS fixed point we conclude that $\Gamma-1$  is 
also a slowly varying function of $N$.

Notice that the  previous features  are not anymore true for $1/4 
<|\beta|<1/2$. This is because in these range of $\beta$ the 
denominator  in Eq. (\ref{gm1})  has a zero at finite negative values 
of  $\phi$, namely for $\cosh\g \phi= -(1+16\b^{2})/(1-16\b^{2})$.

Being $\Gamma$ nearly constant and $\Gamma>1$, we  have a tracker 
behaviour of our solutions till the scalar field  reaches values of 
order $1/\gamma$. In this region we have (see e.g. \cite{amendola})
\beq\lb{u2}
w_{\phi}=\frac{w_{M}-2(\Gamma-1)}{1+2(\Gamma-1)}.
\eeq
Being $w_{\phi}<w_{M}$ dark energy evolves more slowly then matter.
Also $\l$ and the ratio $\Omega_{\phi}/\Omega_{M}$ varies slowly. 
$\l$ decreases toward zero, whereas $\Omega_{\phi}/\Omega_{M}$ 
increases.   
The main difference between our model and the usual tracker solutions 
is the way in which the universe exits the scaling behaviour and   
produces the cosmic acceleration. 

In the usual scenario this happens 
when $\l$ reaches the lower bound for stability of the scaling 
solution,  $\l^{2}=3(1+w_{M})$. 
One can  easily check that for our models this happens instead  when the 
system reaches the region where the approximation of slow varying $\l$ and $\Gamma$
does not hold anymore. The universe exits the scaling regime when it 
reaches the regions $\phi \sim -1/\g$ where $\Gamma$ and $\l$ vary very 
fast and we have a sharp  transition to the dS phase. This transition is the 
cosmological counterpart of the hyperscaling violating/AdS phase 
transition in holographic theories of gravity \cite{Cadoni:2009xm,Gouteraux:2012yr}.

We are now in position of giving a detailed, albeit qualitative, 
description of the global behavior of our FLRW solutions. 
This  behaviour depends on the range of variation of the parameter 
$\beta$. We have to distinguish three different cases: $(I) : 
\b<\b_{1};\, (II): \b_{1}<\b<\b_{0};\, (III): \b>\b_{0}$ with 
$\b_{0,1}$ given by Eq. (\ref{op}) and (\ref{rd}). 

In case $(I)$ the scaling solution, describing the universe at early 
times,  is a stable spiral and $\Omega_{\phi}/\Omega_{M}\approx 1$.
As the cosmic time increases, $\Omega_{\phi}/\Omega_{M}$ stays almost 
constant and  $\l$ 
decreases   toward the value  $\l^{2}=24(1+w_{M})^{2}/(7+9w_{M})$ 
below which the scaling solution is  a stable node. However, this 
value  is not in the region of slow varying $\l$. Cosmological 
evolution undergoes a sharp  transition to  the dS accelerating 
phase. The behaviour of $\l$, 
$\Gamma-1$, $\Omega_{\phi}$ and $w_{\phi}$ as  a function of $\phi$  
for this class of solutions is explained  in Figs. 
\ref{fig:lG}, \ref{fig:wO}, where we 
plot as representative element  
$\b=-15/64$ and we take nonrelativistic matter, $w_{M}=0$.

Notice that  Figs \ref{fig:wO} have been produced using the 
expressions  (\ref{fg}),  (\ref{u2}) respectively for $\Omega_{\phi}$ 
and $w_{\phi}$, which are valid in  the region of slow variation 
of  $\l$ and $\G$. Therefore,  the plots can be trusted  only in the 
region $\phi<<-1/\g$.

In case $(II)$ the scaling solution, describing the universe at early 
times,  is a stable node and $\Omega_{\phi}/\Omega_{M}= {\cal{O}} (1)$ 
but $\Omega_{\phi}>\Omega_{M}$. At early times $\l$ decreases  very 
slowly. As explained above, there is no smooth transition to the 
accelerating scaling phase (\ref{kk})  with $\Omega_{\phi}=1$  
but a sharp transition to the de Sitter phase. The behaviour of $\l$, 
$\Gamma-1$, $\Omega_{\phi}$ and $w_{\phi}$ as  a function of $\phi$  
for this class of solutions is explained  in Figs. \ref{fig:lG}, \ref{fig:wO},
where we 
plot as representative element  
 $\b=-1/8$ and  $w_{M}=0$.

In case $(III)$ the scaling solution is a saddle point and at early 
times the 
accelerating, scalar-field dominated solution (\ref{kk}) is stable.
We have $w_{\phi}<-1/3$ and $\Omega_{\phi}=1$.
Here we  have a transition from a power-law, accelerating universe at 
early times to the de Sitter solution at late times.
Obviously, this case is not realistic because it cannot  
describe  the  matter dominated  era.
The plot of $\l$, 
$\Gamma-1$, $\Omega_{\phi}$ and $w_{\phi}$ as  a function of $\phi$  
for this class of solutions is depicted  respectively in Figs. 
\ref{fig:lG}, \ref{fig:wO} for $\b=0$ and  $w_{M}=0$.
\begin{figure}[ht]
\begin{center}
\begin{tabular}{cc}
\epsfig{file=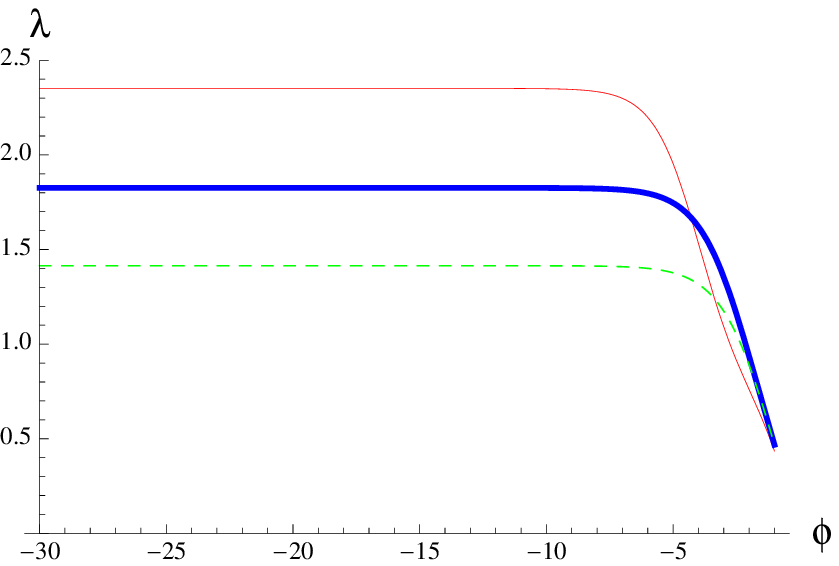,width=7cm,angle=0}&
\quad\epsfig{file=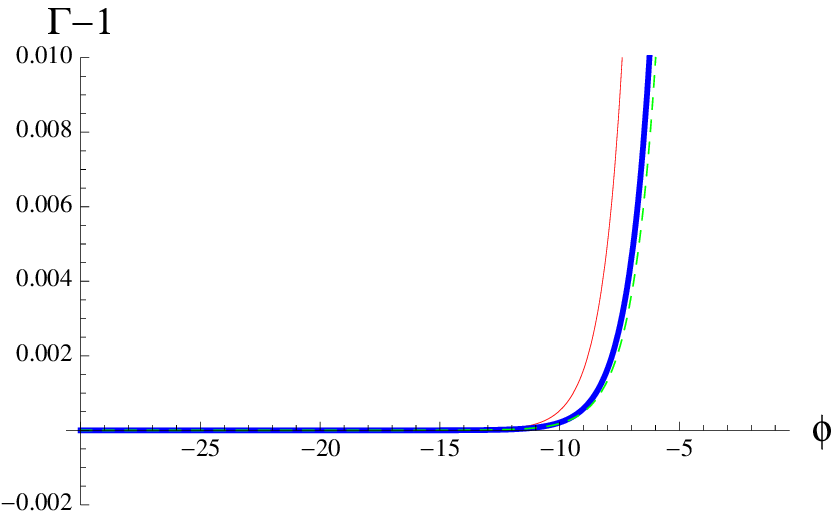,width=7cm,angle=0}
\end{tabular}
\caption{Plot of the function $\lambda(\phi)$ (left panel) and 
$\Gamma(\phi)-1$ (right panel)  for selected values of 
the parameter $\b$   representative of the three classes of solutions 
($I,II$ and $III$) and for $w_{M}=0$. 
The thin, red  lines  are  the plots relative to the model  of class 
$I$  with $\b= -15/64$. The  thick blue lines give the plots of a model 
in class $II$ with  $\b=-1/8$. The green, dashed lines correspond to a 
model in class $III$ with $\b=0$.
\label{fig:lG}}
\end{center}
\end{figure}

\begin{figure}[ht]
\begin{center}
\begin{tabular}{cc}
\epsfig{file=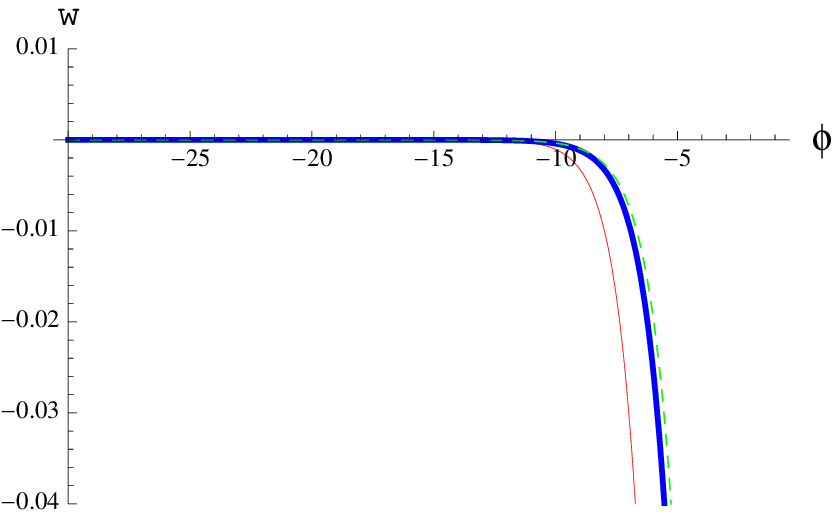,width=7cm,angle=0}&
\quad\epsfig{file=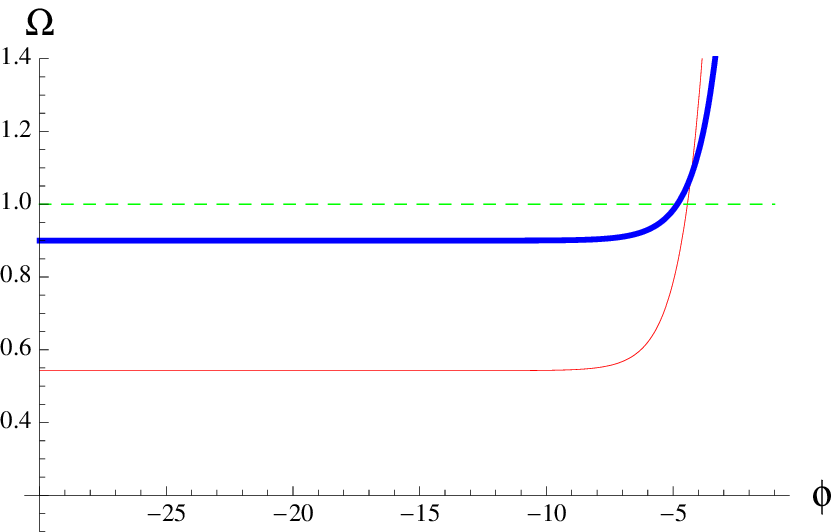,width=7cm,angle=0}
\end{tabular}
\caption{Plot of the function $w_{\phi}(\phi)$ (left panel) and 
$\Omega_{\phi}(\phi)$ (right panel)  for selected values of 
the parameter $\b$   representative of the three classes of solutions 
($I,II$ and $III$) and for $w_{M}=0$. 
The thin, red  lines  are  the plots relative to the model  of class 
$I$  with $\b= -15/64$. The  thick blue lines give the plot of a model 
in class $II$ $\b=-1/8$. The green, dashed lines correspond to a 
model in class $III$ with $\b=0$.
\label{fig:wO}}
\end{center}
\end{figure}

Let us conclude this section with  a brief, general discussion   about  the parameters entering 
in our model. Basically, apart from the Planck mass 
in the action (\ref{action}) enter  a dimensionless parameter $\beta$ and a 
length scale  $R_{0}$ (Notice that in Eq. (\ref{action})  we have set 
$\k^{2}=8\pi G=1/2$). In addition we  have the integration 
constants  of the differential equations (\ref{ds}), which have to be 
determined by solving the Cauchy problem. Some of these 
constants will be related  
to  $t_{-}$ and $a,b$ characterizing respectively   the power-law 
(\ref{sca})  and   
the exponential regime (\ref{sol4}). However, the scale symmetries of the 
gravitational background together with the attractor behaviour  of 
the scaling solution and of the de Sitter fixed point make the 
cosmological dynamics largely, if not completely, independent from 
the initial conditions. Cosmological evolution can be seen as a flow 
from a scaling fixed point  to a conformal dS fixed point, in which 
the system looses any memory about initial conditions.
 The final state is therefore completely characterized by the length 
 scale $R_{0}$, which determines everything (Hubble parameter, 
 acceleration, cosmological constant and the mass for the scalar, see 
 Eq. (\ref{ds1}). The length scale $R_{0}$ can be fixed by the dark 
 energy density necessary to explain  the present  acceleration of 
 the universe,
 $\rho_{de}= 10^{-123}m_{p}^{4}$. This gives a mass of the scalar 
 $m\approx 10^{-33}eV$.
 
 On one side this uniqueness gives a lot of predictive power to the 
 model, but on the other side the presence of an extremely light 
 scalar excitation runs into the the well-known  problems  in the 
 framework of particle physics, SUGRA theories and cosmological 
 constant scenarios \cite{Peebles:2002gy,Padmanabhan:2002ji}.

\section{Conclusions}
In this paper we have shown that scalar solitonic solutions of holographic 
models with hyperscaling 
violation have an interesting cosmological counterpart, which can be 
obtained by analytical continuation and by flipping the sign of the 
potential for the scalar field.  The resulting flat FLRW solutions 
can be used to model cosmological evolution driven by dark energy 
and usual matter. 

In absence of matter, the flow from the hyperscaling 
regime to the conformal 
AdS fixed point in holographic models 
correspond  to cosmological evolution from  power-law regime at early 
cosmic times    to  a 
dS fixed point at late times. 
 In presence of matter, we have a 
scaling regime at early times, followed by an intermediate regime 
with tracking behaviour. At late times the solution exits the 
scaling regime  with a sharp  transition to a  de 
Sitter spacetime. The phase 
transition between hyperscaling violation and conformal fixed point 
observed in holographic gravity has a cosmological analogue in  the 
transition between a scaling,  era  and  a dS era 
dominated by the energy of the vacuum.

We have been able to solve exactly the dynamics only in absence of 
matter. When matter is present we  do not have full control of the 
global solutions. Nevertheless, by writing the  
cosmological equations as a dynamical system and by investigating 
three approximated regimes we have given strong evidence that the 
above  
picture  is   realized.   

At the present stage our model for dark energy cannot be completely 
realistic. In the matter dominated epoch the ratio 
$\Omega_{\phi}/\Omega_{m}\approx 1$, so that we have a problem with 
nucleosynthesis.
Moreover, the late-time cosmology shares the same problems of 
 all cosmological  constant scenarios. The vacuum energy is an 
unnaturally tiny free parameter of the model. The same is true for 
the mass of the scalar excitation associated to the quintessence 
field.

There are several open questions, which are worth to be investigated 
in order to support the above picture.
One should derive the exact full phase space description of the dynamical 
system in presence of matter 
to check the correctness of our results.
In particular, having full control on the  phase space would 
give a precise description of the sharp transition  between the scaling 
and  the dS regime. This would also help us to shed light on the analogy 
between the  cosmological transition and the hyperscaling violation/ 
AdS holographic phase transition.

Other key points that could  improve our knowledge on the subject 
are: (1) Comparison between the cosmological dynamics and the RN group 
equations for the holographic gravity theory; (2) understanding 
of the analogy phase transition/cosmological transition in terms of 
the analytical continuation.   

\begin{acknowledgements}
We thank O. Bertolami and S. Mignemi for discussions and valuable 
comments.
\end{acknowledgements}

\bigskip

\bigskip

\bigskip

\bigskip

\bigskip

\bigskip

\bigskip

\bigskip

\bibliography{darkenergy}

\end{document}